\documentclass[letterpaper]{article}
\usepackage[T1]{fontenc}
\usepackage{geometry}
\usepackage{braket}
\usepackage{setspace}
\usepackage{titletoc}
\usepackage{siunitx}
\DeclareSIUnit\angstrom{\text {Å}}

\usepackage[
    sorting=none,
    url=false,
    isbn=false,
    eprint=true,
    doi=true
]{biblatex}
\AtEveryBibitem{%
  \clearfield{langid}
  \clearlist{language}
  \clearfield{note}
}

\usepackage{graphicx}
\usepackage{float}
\usepackage{appendix}
\newfloat{scheme}{htbp}{los}
\floatname{scheme}{Scheme}
\floatname{chart}{Chart}
\newfloat{graph}{htbp}{loh}
\usepackage{chemformula}
\usepackage[version = 4]{mhchem}
\usepackage{siunitx}
\usepackage{hyperref}
\hypersetup{hidelinks,
colorlinks,
linkcolor={red!40!black},
    citecolor={blue!50!black},
    urlcolor={blue!70!black}}
\usepackage[nameinlink]{cleveref}
\usepackage{authblk}
\author[1$^\S$]{Fabian A. Gerritsma}
\author[1$^\S$]{Roland W. Mulder}
\author[2]{Balázs Nagyfalusi}
\author[1]{Pieter C.J.A. van Zweeden}
\author[1$^\#$]{Samer Kurdi}
\author[3]{Francisco Machado}
\author[2]{Jaime Ferrer}
\author[1,4]{Samuel Mañas-Valero*}
\author[1]{Toeno van der Sar**}
\affil[1]{Department of Quantum Nanoscience, Kavli Institute of Nanoscience, Delft University of Technology, 2628CJ Delft, The Netherlands}
\affil[2]{Departamento de Física, Universidad de Oviedo, 33007 Oviedo, Spain}
\affil[3]{QuTech, Kavli Institute of Nanoscience, Delft University of Technology, 2628CJ Delft, The Netherlands}
\affil[4]{Instituto de Ciencia Molecular (ICMol), Universitat De València, 46980 Paterna, Spain}

\title{Thermally Activated Magnetic Switching in Antiferromagnetic Van der Waals Trilayers Resolved by Single-Spin Diamond Magnetometry}
\date{$^\S$F.G. and R.M. contributed equally to this work\\
*Email: samuel.manas@uv.es\\
**Email: t.vandersar@tudelft.nl}

\begin{document}
\begin{refsection}
\maketitle

\begin{abstract}
    Van der Waals magnets provide a versatile platform for exploring magnetic phenomena that emerge from reduced dimensionality, including layer-dependent magnetic ordering. Here, we use scanning nitrogen-vacancy (NV) centers in diamond magnetometry to measure the magnetization of a trilayer of the A-type antiferromagnet CrPS$_4$, resolving its zero-field bistability between the up-down-up ($\uparrow\downarrow\uparrow$) and down-up-down ($\downarrow\uparrow\downarrow$) layer magnetization configurations. The coercive field decreases approximately linearly with temperature, consistent with the N\'eel--Arrhenius model. Density Functional Theory calculations capture the temperature dependence of the magnetization and determine the coercive field in the coherent-rotation case. The measured coercive fields are lower, identifying thermally activated domain nucleation as the reversal mechanism. Spatial magnetic imaging supports this picture, revealing independently switching regions likely pinned by structural inhomogeneities. Our results highlight the power of scanning NV magnetometry for the quantitative investigation of magnetization reversal in two-dimensional magnets, unraveling the magnetic phenomenology that emerges in low-dimensional systems.
\end{abstract}

\section*{Keywords}

van der Waals magnet, antiferromagnet, magnetization reversal, nitrogen vacancy center

\section{Introduction}
Van der Waals magnets provide a unique platform for investigating low-dimensional magnetism and tailoring magnetic functionality at the atomic scale, bridging fundamental studies of correlated spin phenomena with emerging spintronic and magnonic applications \cite{burchMagnetismTwodimensionalVan2018,kurebayashiMagnetismSymmetrySpin2022,ahnProgressProspectsTwodimensional2024,manas-valeroFundamentalsApplicationsVan2025}. Thinning down van der Waals magnets to the two-dimensional (2D) limit can profoundly alter their magnetic properties, giving rise to phenomena such as layer-dependent magnetic ordering \cite{huangLayerDependentMagnetismSpin2023a}, thickness-dependent magnon spectra \cite{myhreThicknessdependentMagnonSpin2026} or lateral exchange bias \cite{wangConfigurableAntiferromagneticDomains2025,pellet-maryLateralExchangeBias2025}.

A paradigmatic manifestation of reduced dimensionality occurs in antiferromagnetic magnetization reversal, which is governed by the competition between magneto-crystalline anisotropy, exchange, demagnetization and Zeeman energies \cite{gharaMagnetizationReversalAntiferromagnetic2023}. The interplay between these energies determines the stabilization and dynamics of spin textures such as domain walls, vortices and coherently rotating states, and governs switching processes driven by magnetic fields, electrical currents, and ultrafast optical excitations \cite{hellmanInterfaceinducedPhenomenaMagnetism2017}. Magnetization reversal in antiferromagnets is of particular interest due to its implications for ultrafast and densely integrated spintronic technologies such as information storage \cite{baltzAntiferromagneticSpintronics2018}.

In layered antiferromagnets, the interlayer exchange coupling and its evolution with thickness govern both the magnetic ground state and its switching \cite{xianSpinMappingIntralayer2022,guoLayerdependentSpinorbitTorque2025,wangFerromagnetlikeBinarySwitching2026}. A-type antiferromagnets such as CrPS$_4$ combine ferromagnetic intralayer order with antiferromagnetic interlayer coupling (\cref{fig:fig1}a), so their magnetic properties depend strongly on layer number and, in particular, on its parity \cite{wangConfigurableAntiferromagneticDomains2025,wangFerromagnetlikeBinarySwitching2026}. In atomically thin odd-layer stacks, the Zeeman energy of the uncompensated layer can overcome the anisotropy barrier of the entire stack, whose magnitude grows with the number of layers \cite{wangConfigurableAntiferromagneticDomains2025}. This enables reversal of the entire stack (from $\downarrow\uparrow\downarrow$ to $\uparrow\downarrow\uparrow$, where the arrows indicate the layer's magnetization) at fields well below the spin-flop transition. This process, however, is inaccessible in bulk, where the spin-flop occurs first \cite{wangDeterminingPhaseDiagram2019,tianLayerdependentMetamagneticTransitions2025}.

Single-spin diamond magnetometry employs the Nitrogen Vacancy (NV) defect in diamond as a magnetic sensor with a spatial resolution in the tens to hundreds of nm range, and static magnetic field sensitivity below \qty{1}{\micro\tesla\per\sqrt{\hertz}} \cite{schirhaglNitrogenVacancyCentersDiamond2014}. The combination of high magnetic field sensitivity and spatial resolution enables the determination of the magnetization of atomically thin van der Waals magnets \cite{thielProbingMagnetism2D2019,zhangAcSusceptometry2D2021,mclaughlinQuantumImagingMagnetic2022,liMultimodalScanningprobeQuantum2026}.

Here, we use scanning NV magnetometry to quantify the magnetization of a CrPS$_4$ trilayer as a function of applied magnetic field and temperature. By using an out-of-plane NV tip, we are able to characterize the flake's magnetization in out-of-plane bias fields that reach up to \qty{0.8}{\tesla}, which are large fields for NV magnetometry experiments \cite{dingDirectImagingMagnetotransport2026}, without loss of imaging quality. We resolve the two bistable magnetic states of the trilayer at zero field: $\downarrow \uparrow \downarrow$ and $\uparrow \downarrow \uparrow$. Our Density Functional Theory (DFT) calculations reproduce the temperature dependence of the magnetization and determine the coercive field expected for coherent rotation. The measured coercive fields are far lower and decrease approximately linearly with temperature, as described by the N\'eel--Arrhenius model, identifying thermally activated domain nucleation as the reversal mechanism. Magnetic imaging confirms this picture by directly visualizing independently switching domains. Our results reveal magnetization-reversal processes that become directly accessible for atomically thin antiferromagnets and demonstrate the power of scanning NV magnetometry for the quantitative study of nanoscale magnetism.

\section{Results and discussion}
\begin{figure}
    \centering
    \includegraphics{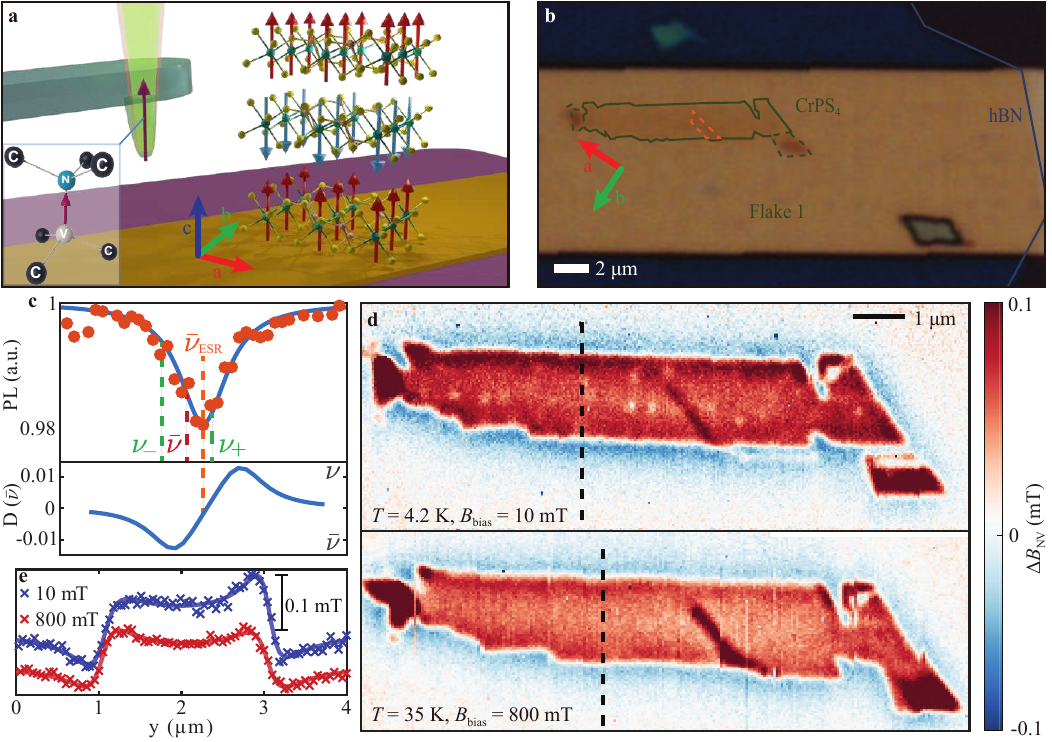}
    \caption{\textbf{Imaging an antiferromagnetic CrPS$_4$ trilayer using NV magnetometry.} (a) Measurement schematics: a CrPS$_4$ trilayer (right) is placed on top of a gold stripline and imaged using an NV scanning tip (left). For CrPS$_4$, green, yellow and pink balls represent Cr, S and P atoms, respectively, where out-of-plane antiparallel red and blue arrows denote the layer's magnetization in the ordered state. The scanning tip (left) contains a single NV center, with green, blue, and gray balls representing carbon atoms, the nitrogen atom, and the vacancy, respectively. The NV center has an effective spin \(S=1\) oriented out of plane (purple arrow) and is excited by a green laser, with its photoluminescence (PL) collected. (b) Optical image of a CrPS$_4$ trilayer (flake 1), outlined in green, stamped onto a coplanar waveguide and encapsulated by h-BN. The outer ends of the flake are bulk regions, outlined by green dashed lines. There is a small region where the flake shows a fold, outline by the orange dashed line. (c) Bimodal feedback algorithm for time-optimized NV magnetometry \cite{navaantonioMagneticImagingStatistical2021}. In the first pixel of each line scan, a full electron spin resonance (ESR) spectrum is acquired by sweeping the microwave frequency while monitoring the photoluminescence (PL) (top graph). The ESR frequency, $\nu_{\mathrm{ESR}}$, corresponding to the minimum of the PL signal, is used as the initial estimate for the feedback algorithm. The bottom graph shows the error function $D(\bar{\nu})$, for which $D = 0$ when the estimate $\bar{\nu} = \nu_{\mathrm{ESR}}$. During the feedback procedure, only two microwave frequencies, $\nu_{\pm}$, applied symmetrically about $\bar{\nu}$, are required to update the estimate by minimizing $|D(\bar{\nu})|$. Further details are provided in Supporting Information \cref{sec:bimodelreference}. (d) 2D magnetic field scans of flake 1 at $B_{\text{bias}}$ and $T$ indicated in the figures. We observe a positive stray field above the flake, indicating a net magnetization aligned with the external field and the NV axis. (e) Line scans taken at the location of the dashed black lines in (d). The blue and red curves show fits to two Lorentzian and two odd-Lorentzian functions, with each Lorentzian pair corresponding to one edge of the flake. Further details of the fitting procedure are provided in Supporting Information \cref{sec:fitting}. The asymmetry of the blue curve is a result of the in-plane magnetization component of the sample.}
    \label{fig:fig1}
\end{figure}

CrPS$_4$ is a van der Waals A-type antiferromagnet (AFM), characterized by ferromagnetic intralayer coupling and antiferromagnetic interlayer coupling below the Néel temperature ($T_\text{N} \sim $ 36-\qty{40}{\kelvin} in bulk crystals \cite{peiSpinDynamicsElectronic2016,budkoMagneticOrderVan2021,guPhotoluminescentQuantumInterference2020,calderMagneticStructureExchange2020}), as shown in \cref{fig:fig1}a. It has a monoclinic crystal structure (space group C2) with a monoclinic angle $\beta \sim \qty{92}{\degree}$ \cite{calderMagneticStructureExchange2020}, making it nearly orthorhombic. The magnetic easy ($c$-)axis lies nearly perpendicular to the layers along the $ab$-plane, with reported canting angles of the magnetic moment of \qty{9.5}{\degree} to \qty{12}{\degree} towards the intermediate ($a$-)axis \cite{calderMagneticStructureExchange2020,pengMagneticStructureMetamagnetic2020,baralCorrelationDrivenSpinReorientation2026}. Its van der Waals nature allows for mechanical exfoliation down to a single layer \cite{houExtraordinaryMagneticSecond2024}. In \cref{fig:fig1}b, we show flake 1, which is a trilayer encapsulated by hexagonal boron nitride (hBN). Crystal growth and fabrication details are given in Supporting Information \cref{sec:sampleprep}. The crystal axes are assigned based on the preferential cleavage directions reported for CrPS$_4$ \cite{leeStructuralOpticalProperties2017}. In a trilayer, the antiferromagnetic interlayer coupling leaves one layer uncompensated, giving two degenerate zero-field ground states, $\uparrow\downarrow\uparrow$ and $\downarrow\uparrow\downarrow$. These states are related by time reversal and carry opposite net magnetization, which can be controlled by  the sign and amplitude of the previously applied magnetic field.

We measure the stray magnetic field of flake 1 as a function of temperature, $T$, and bias magnetic field $B_{\text{bias}}$ in the $\hat{z}$-direction. We make use of a scanning tip with an out-of-plane (111) oriented single NV center, as sketched in \cref{fig:fig1}a. The component of the stray magnetic field of the flake along the axis of the NV center causes a Zeeman splitting of the electron spin resonance (ESR) (\cref{fig:fig1}c) which is detected via the optical detected magnetic resonance (ODMR). By monitoring this splitting, the magnitude of the magnetic field is found. To optimize the scanning speed, during a line scan, the resonance is tracked using the bimodal feedback scheme shown in \cref{fig:fig1}c \cite{navaantonioMagneticImagingStatistical2021}. Further details are provided in Supporting Information \cref{sec:fitting}.

Scanning NV magnetometry has a wide temperature and magnetic field range accessible, as shown in \cref{fig:fig1}d, which displays two-dimensional (2D) magnetic field images acquired at different temperatures and bias magnetic fields. In Fig. \cref{fig:fig1}d we show two representative magnetic images. The top panel is obtained in the low-temperature antiferromagnetically ordered state (\qty{4.2}{\kelvin}, \qty{10}{\milli\tesla}), where the magnetic contrast corresponds to the one-layer uncompensated magnetization. The bottom panel corresponds to close to $T_N$ (\qty{35}{\kelvin}, \qty{0.8}{\tesla}), where the trilayer is in the field-induced canted state. By performing analogous measurements at different temperatures and magnetic fields, we have experimental access to the magnetic phase diagram of the CrPS$_4$ trilayer.

To find the magnetization of the flake, line scans are acquired along the trajectory indicated in \cref{fig:fig1}d while maintaining a constant tip--sample lift height (see Supporting Information \cref{sec:bimodelreference}). Example line scans are shown in \cref{fig:fig1}e. We extract the two-dimensional magnetization by fitting the measured stray-field profiles to the analytical expression for the stray field of a homogeneously magnetized two-dimensional flake. It consists of a linear combination of a Lorentzian and its odd counterpart, coming from the magnetization perpendicular to and along the axis of the NV center, respectively, at each edge. The low-temperature example in \cref{fig:fig1}e at \qty{10}{\milli\tesla} leads to $|\mathbf{M}| = \qty{0.136(8)}{\milli\ampere}$, consistent with the monolayer magnetization reported in literature \cite{calderMagneticStructureExchange2020}. Further details of the fitting procedure are provided in
Supporting Information \cref{sec:fitting}.

\begin{figure}
    \centering
    \includegraphics{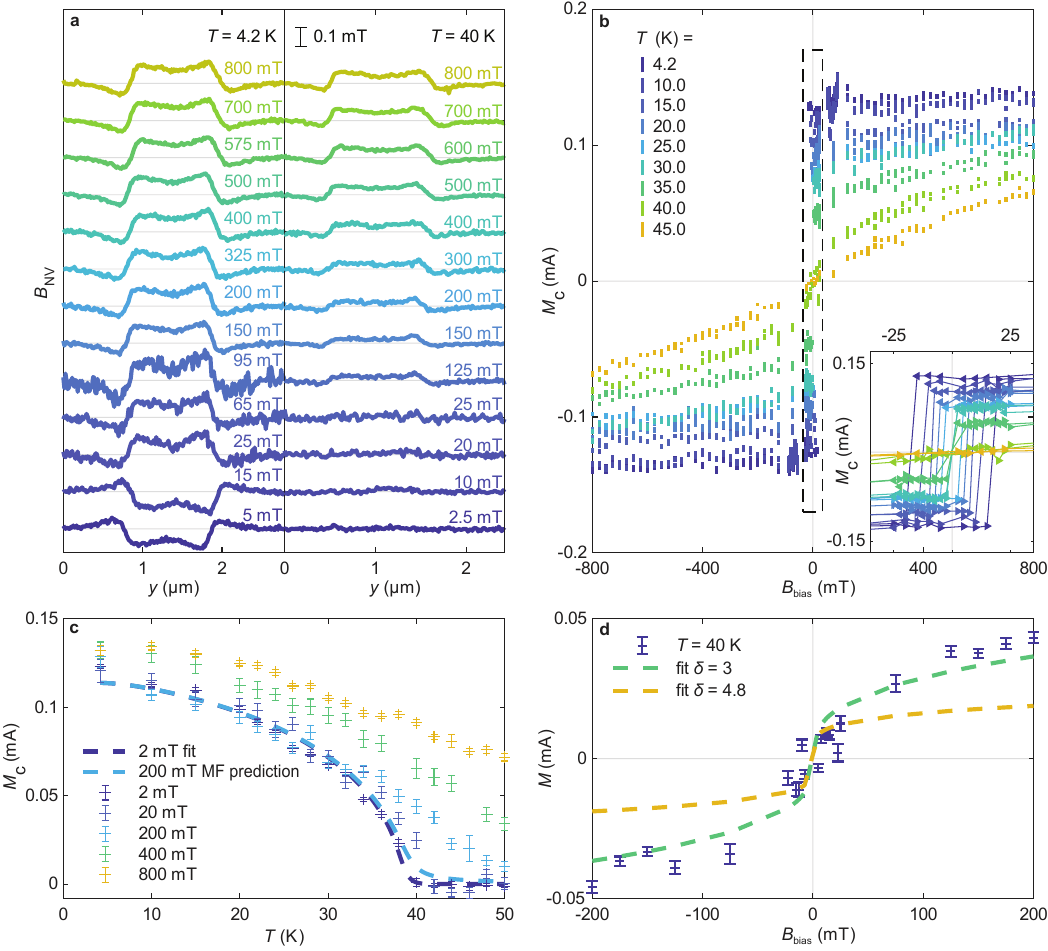}
    \caption{\textbf{Magnetization of CrPS$_4$ as a function of $B_{\text{\normalfont{bias}}}$ and $T$.} (a) Line scans of flake 1 during an upward magnetic field sweep for $T = \qty{4.2}{\kelvin}$ (left) and $T = \qty{40}{\kelvin}$ (right). (b) Magnetization along the $c$-axis as a function of $B_{\text{bias}}$ during upward and downward field sweeps. The inset shows a zoomed view of the low-field data, revealing a hysteresis loop. Triangles indicate the sweep direction. Around $B_{\text{bias}} = \qty{100}{\milli\tesla}$, the NV center exhibits a reduced sensitivity due to the ground-state level anticrossing of its spin levels \cite{happacherTemperatureDependentPhotophysicsSingle2023}. (c) $c$-component of the magnetization as a function of temperature during an upward temperature sweep. The dashed lines show calculations based on a mean-field model for an antiferromagnetic trilayer. The \qty{2}{\milli\tesla} experimental data are used to fit for the material dependent constants in our mean-field model; the resulting \qty{200}{\milli\tesla} curve is given by using the same constants and only modifying the applied field. (d) Magnetization from (b) as a function of $B_{\text{bias}}$ at $T = T_\text{N} \sim $ \qty{40}{\kelvin}, truncated at 30\% of its saturation value. The dashed lines show fits to $m \propto h^{1/\delta}$.}
    \label{fig:fig2}
\end{figure}

We acquire line scans of flake 1 as a function of bias magnetic field for different fixed temperatures, revealing magnetization mechanics such as magnetization reversal. \Cref{fig:fig2}a shows representative stray-field line scans of flake 1 acquired at $T = \qty{4.2}{\kelvin}$ and $T \sim T_\text{N} = \qty{40}{\kelvin}$ for a range of bias magnetic fields during an upward sweep after saturating at \qty{-1}{\tesla}. At \qty{4.2}{\kelvin}, we observe a magnetization reversal between \qty{15}{\milli\tesla} and \qty{25}{\milli\tesla}, switching from the $\downarrow \uparrow \downarrow$, stabilized after the initial saturation at large negative fields, to the $\uparrow \downarrow \uparrow$ state. Due to the low temperature, the flake is stabilized in the $\uparrow \downarrow \uparrow$ state for our accessible bias magnetic field range up to \qty{800}{\milli\tesla}. At $T = \qty{40}{\kelvin}$, the trilayer is in the paramagnetic phase, so it does not have spontaneous  magnetization. Therefore, near zero field, the magnetization, and hence the stray-field signal, vanishes and no magnetic reversal is observed. An applied field induces a magnetization that grows rapidly with field, because the susceptibility is strongly enhanced close to $T_{\text{N}}$ \cite{blundellMagnetismCondensedMatter2014}, and the stray fields are sensed by the NV center.

With increasing temperature, thermally excited spin waves progressively reduce the magnetization, which vanishes at the Néel temperature as the trilayer becomes paramagnetic. This is visible in \cref{fig:fig2}b,c, where we extract the magnetization from line scans such as in \cref{fig:fig2}a. \Cref{fig:fig2}b shows the field-dependence of the extracted magnetization along the $c$-axis. The magnetization along the $a$-axis is shown in Supporting Information \cref{fig:magXvsB}. From these results, we determine the low-temperature canting angle of the crystal to be $\sim \qty{20(3)}{\degree}$. This value is much larger than the \qty{9.5}{\degree} to \qty{12}{\degree} canting found by neutron scattering in bulk CrPS$_4$ \cite{calderMagneticStructureExchange2020,pengMagneticStructureMetamagnetic2020,baralCorrelationDrivenSpinReorientation2026}, and in agreement with the larger ($\sim$\qty{16}{\degree}) values observed in five-layer CrPS$_4$ by scanning NV measurements. 

A spin-flop transition for bulk CrPS$_4$ under out-of-plane field was previously measured at \qtyrange{0.7}{0.8}{\tesla} \cite{calderMagneticStructureExchange2020,pengMagneticStructureMetamagnetic2020}. For a trilayer, the transition is expected to occur at higher field due to the Zeeman energy of the uncompensated layer, which lowers the energy of the antiferromagnetic ground state, thereby stabilizing it \cite{wangDeterminingPhaseDiagram2019,tianLayerdependentMetamagneticTransitions2025}, as we develop below using DFT. We do not observe any indication of such a transition within our accessible field range, suggesting that the spin-flop field is indeed higher than \qty{0.8}{\tesla}.

At low bias fields, a hysteresis loop is visible, showing the bistable $\uparrow \downarrow \uparrow$ and $\downarrow \uparrow \downarrow$ states. To characterize this, we increase the measurement density for bias magnetic fields below \qty{25}{\milli\tesla}. For clarity, this low-field region is shown in the inset, where the triangles denote the data points and indicate the sweep direction. The hysteresis loop gradually decreases with increasing temperature and disappears above $T_\text{N}$, displaying tunability of the coercive fields.

We can describe the thermal dependence of the magnetization for the full temperature range with a mean-field model. To find this relation, we also sweep the temperature while keeping the bias magnetic field fixed, as shown in \cref{fig:fig2}c. We employ a mean-field model for an A-type antiferromagnetic trilayer to describe this temperature dependence. The model is based on a spin Hamiltonian with ferromagnetic intralayer and antiferromagnetic interlayer exchange interactions in the presence of an applied magnetic field (full derivation is provided in Supporting Information \cref{sec:MFT}) \cite{serhaUnifiedTheoryMagnetization2025}. Within the mean-field approximation, the outer two layers are assumed to be equivalent, reducing the problem to two coupled self-consistency equations for the layer magnetizations. Solving these equations yields the total magnetization as a function of temperature and magnetic field.

The mean-field model reproduces the overall decrease of the magnetization with increasing temperature and captures the transition to the paramagnetic phase near $T_\mathrm{N}$. In \cref{fig:fig2}c, the mean-field fit to the \qty{2}{\milli\tesla} data is shown as a dashed curve. The dashed curve for \qty{200}{\milli\tesla} is calculated using the same fitted parameters, with only the applied magnetic field changed. While the model correctly predicts that the transition smears to higher temperature with increasing field, the magnitude is underestimated. This discrepancy likely reflects the simplifying assumptions of the mean-field approximation. Deviations close to $T_\text{N}$ are expected because the mean-field approximation neglects critical fluctuations, while the low-temperature deviations arise because mean-field theory does not account for the reduction of the magnetization by thermal spin-wave excitations.

Near the Néel temperature, the free energy can be expanded as: \cite{landauStatisticalPhysics1938}
\begin{equation}
    F \sim am^2 + bm^4 - hm,
\end{equation}
where $m$ is the magnetization, $a = \alpha(T - T_\text{N})$ since it changes sign at $T_\text{N}$, $h$ is the bias field, and $\alpha, b > 0$ are material-dependent constants in arbitrary units. The magnetization is captured by the minimum of the free energy, so where
\begin{equation}\label{eq:freeeng}
    \frac{\partial F}{\partial m} = m(2a + 4bm^2) - h = 0.
\end{equation}
Above $T_\text{N}$, the free energy has a single minimum. Below $T_\text{N}$, two minima appear, corresponding to $\uparrow\downarrow\uparrow$ and $\downarrow\uparrow\downarrow$, each scaling as $\sqrt{|T - T_\text{N}|}$, with the lower-energy minimum corresponding to the physically stable solution. At $T_\text{N}$, the low-field magnetization is expected to follow the scaling $m \propto h^{1/\delta}$. Whereas the Landau mean-field theory predicts $\delta=3$, the 2D Ising universality class is characterized by $\delta = 15$ and the 3D Ising or Heisenberg classes by $\delta \approx 4.8$\cite{nishimoriElementsPhaseTransitions2011}. In figure \cref{fig:fig2}d, the $T \approx T_{\mathrm{N}}$ magnetization is compared against two fits with $\delta = 3$ and 4.8. The fits show that the trilayer deviates strongly from the Ising universality class, and is better described by the mean field exponent. This may suggest a contribution from long-range dipolar interactions.

\begin{figure}
    \centering
    \includegraphics{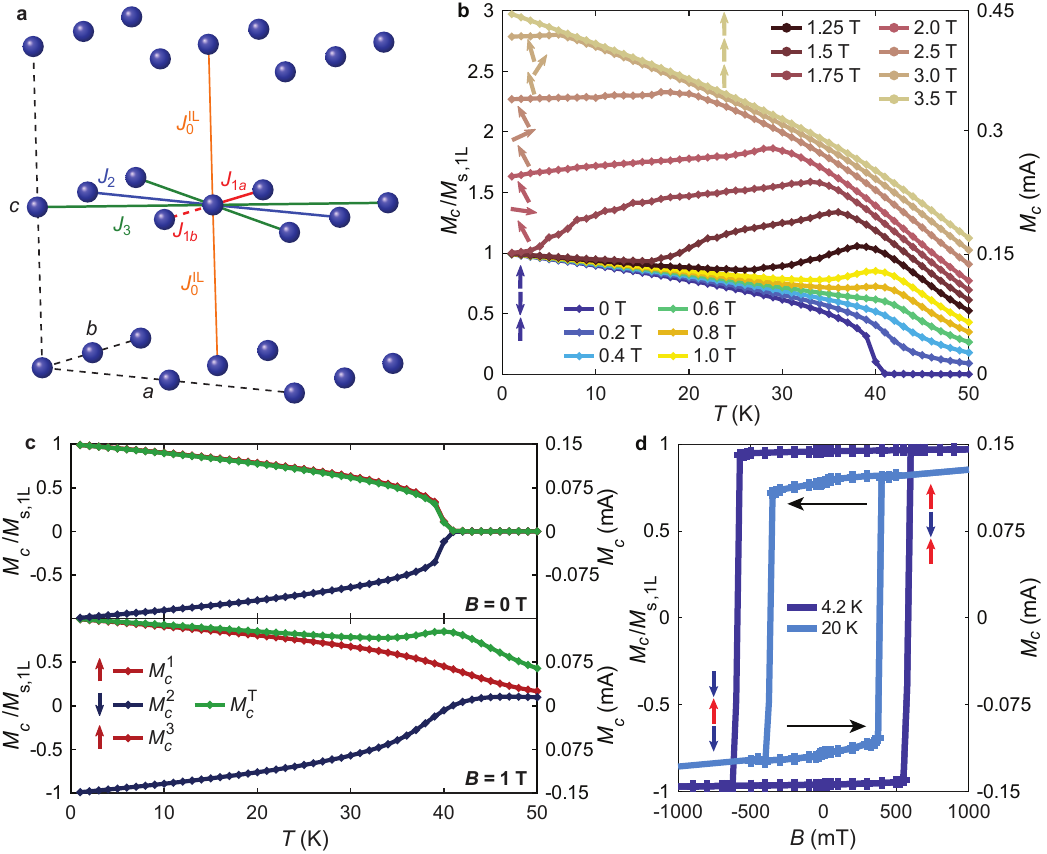}
    \caption{\textbf{DFT-based Monte Carlo simulations on the magnetization of CrPS$_4$.} (a). Visual representation of the considered magnetic interactions in the spin Hamiltonian describing the magnetic CrPS$_4$ trilayer. The different colors denote different interacting shells. The dashed black lines show the crystallographic $a$, $b$, and $c$ directions of the lattice. (b) Magnetization along the $c$-axis of the crystal normalized to the saturation magnetization $M_{\rm{s}}$ of one CrPS$_4$ layer, as a function of temperature for selected external fields. The arrows indicate the layer-resolved magnetization directions in the crystallographic $ac$-plane at low temperature, color matched to the corresponding bias field in the legend. (c) Layer-resolved magnetization along the $c$-axis at $B=0$ and $\qty{1}{\tesla}$, and the sum of the separate magnetization vectors $M_c^\mathrm{T}$. The outer layers, indicated by the same color, have the same magnetization. (d) Calculated hysteresis for a coherent reversal spin switching from $\uparrow \downarrow \uparrow$ to $\downarrow \uparrow \downarrow$, showing zero-field bistability, at \qty{4.2}{\kelvin} and \qty{20}{\kelvin}.
    }
    \label{fig:fig3}
\end{figure}

To model the magnetic response of the trilayer beyond the mean-field approach previously discussed, we first calculate the electronic structure of CrPS$_4$ using the SIESTA density functional theory (DFT) code \cite{Soler2002,Cuadrado2012}, as detailed in Supporting Information  \cref{sec:DFTsup}. We obtain a magnetic moment of 2.943~$\mu_\mathrm{B}$ per Cr site, in agreement with our experimental observations (see Supporting Information \cref{sec:estCr}). To describe the magnetic interactions, we then calculate the intralayer exchange interactions with the GROGU code \cite{Grogu2023}, considering the first three interacting shells, as illustrated in \cref{fig:fig3}a.

The effective interlayer interaction, $J^{\mathrm{IL}}_{0}$, is treated as a fitting parameter, which identifies it to be \qty{0.2}{\milli\electronvolt}. The resulting exchange parameters and anisotropy constants are summarized in \cref{tab:exchange} (Supporting Information \cref{sec:DFTsup}).

From the DFT results, we model the temperature dependence of the magnetization using Monte Carlo (MC) simulations, initialized with a random spin configuration and subsequently cooled in temperature steps of \qty{1}{\kelvin} (see Supporting Information \cref{sec:DFTsup} for further details). The calculated magnetization curves, presented in \cref{fig:fig3}b, reproduce our experimental observations (\cref{fig:fig2}c). At zero field, the magnetization increases sharply below \qty{40}{\kelvin} and saturates at low temperature to the magnetization of a single layer. Under an applied field, the zero-field sharp transition around \qty{40}{\kelvin} broadens and develops a kink at higher fields (see, for example, \qty{0.8}{\tesla}). The calculations also allow us to access fields beyond our experimental capabilities, revealing a transition at low temperatures from a parallel ferromagnetic state ($\uparrow \uparrow \uparrow$) for fields above \qty{3.5}{\tesla}, with magnetization $3M_{\rm{s,1L}}$, to an antiparallel state ($\uparrow \downarrow \uparrow$) below \qty{1.5}{\tesla}, with net magnetization $M_{\rm{s,1L}}$. $M_{\rm{s,1L}}$ is the saturation magnetization of a single layer. At intermediate fields, we obtain a spin-flop state, as sketched by the arrows configurations in \cref{fig:fig3}b.

For a better understanding of the magnetization directions of each layer when an out-of-plane magnetic field is applied, we consider the layer-resolved magnetization curves shown in \cref{fig:fig3}c for $B = 0$ and \qty{1}{\tesla}. In zero field, the magnetization of each layer averages to zero above the critical temperature, while antiferromagnetic (AFM) ordering develops immediately below $T_\mathrm{N}$ ($\uparrow \downarrow \uparrow$). Under an applied field, the magnetic moments already align with the external field above $T_\mathrm{N}$, resulting in a region where all three layers are parallel ($\uparrow \uparrow \uparrow$). Upon further cooling, the magnetization of the middle layer changes sign while the top and bottom layers remain aligned with each other ($\uparrow \downarrow \uparrow$), leading to a decrease in the total magnetization. With increasing field, the reversal of the middle layer occurs at lower temperatures, extending the $\uparrow \uparrow \uparrow$ region and producing a more pronounced local maximum in the $M(T)$ curve.

Finally, we consider the spin-reversal mechanism where all spins in the trilayer rotate coherently, as shown in \cref{fig:fig3}d. The simulation yields a clear hysteresis around zero field, characterized by coercive fields ($B_{\text{co}}$) on the order of \qty{600(25)}{\milli\tesla} at $\qty{4.2}{\kelvin}$ and \qty{375(25)}{\milli\tesla} at \qty{20}{\kelvin}.

\begin{figure}
    \centering
    \includegraphics{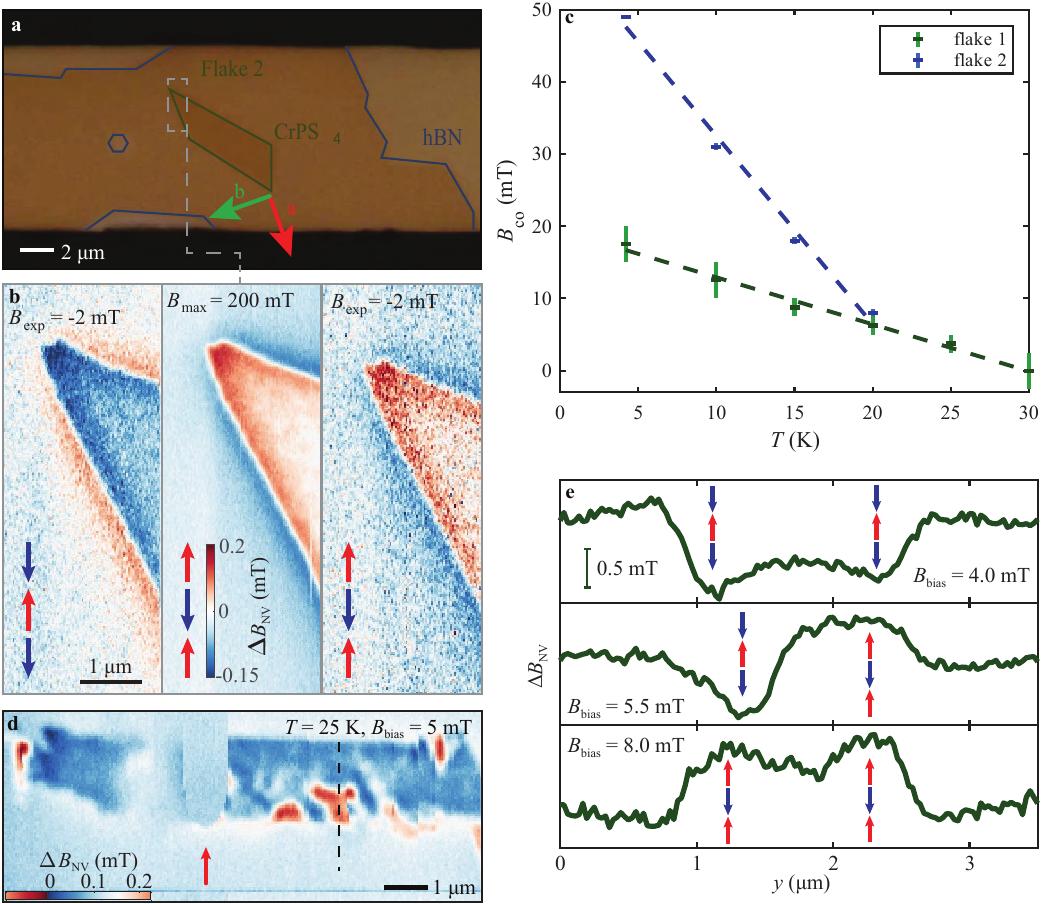}
    \caption{\textbf{Coercive field measurements of magnetization reversal in trilayer CrPS$_4$.} (a) Optical image of a CrPS$_4$ trilayer (flake 2), outlined in green, stamped onto a coplanar waveguide and encapsulated by h-BN. (b) Magnetization reversal of flake 1 at $T = \qty{10}{\kelvin}$. At $B_{\text{bias}} = \qty{-2}{\milli\tesla}$, the stray field points along $-\hat{z}$, corresponding to the $\downarrow \uparrow \downarrow$ configuration. Applying a sufficiently large magnetic field in the $+\hat{z}$-direction, here $B_{\text{bias}} = \qty{200}{\milli\tesla}$, reverses the magnetization to the $\uparrow \downarrow \uparrow$ configuration. Returning to $B_{\text{bias}} = \qty{-2}{\milli\tesla}$ leaves the flake in the $\uparrow \downarrow \uparrow$ state, despite the $\downarrow \uparrow \downarrow$ state being energetically favorable. (c) Coercive fields of both flakes as a function of temperature. The dashed lines show linear fits to the data. (d) Magnetic field image of the damaged flake 1 acquired at a bias magnetic field just below the coercive field of the flake, showing several regions with reversed magnetization. The red arrow indicates a location where damage to the flake occurred. In this image, the damage caused the tip to retract to a distance of $\sim\qty{2}{\micro\metre}$, and subsequently being too far away to detect the stray field from the flake. Following the damage, the line scans were acquired along the dashed line. (e) Line scans acquired at the post-damage line-scan location at $T = \qty{25}{\kelvin}$ during an upward field sweep. At \qty{55}{\milli\tesla}, one region reverses its magnetization. Increasing the field to \qty{80}{\milli\tesla} causes the second region to reverse.}
    \label{fig:fig4}
\end{figure}

The hysteresis calculated for a coherent reversal process yields coercive fields much larger than those observed experimentally (\cref{fig:fig2}b), suggesting a different underlying spin reversal mechanism. To verify this, we introduce a second isolated CrPS$_4$ trilayer (flake 2), shown in \cref{fig:fig4}a, to characterize its magnetization reversal. We determine the $B_{\text{co}}$ of both flakes using two different methods. For flake 1, the ODMR contrast of the NV center remains sufficiently high over the entire applied field and temperature range, allowing direct measurement of the hysteresis loops (inset of \cref{fig:fig2}b) and hence of $B_{\text{co}}$.

For flake 2, however, the NV tip exhibits low ODMR contrast and, consequently, low magnetic-field sensitivity over a wide range of bias fields. Therefore, we determine the coercive field with the first-order-reversal-curve (FORC) protocol, in which the magnetization is always imaged at a fixed field of $B_{\text{exp}} = \qty{-2}{\milli\tesla}$, as illustrated in Supporting Information \cref{fig:coercMethodsFlake2}. The FORC protocol we apply for finding the coercive fields is as follows. First, a magnetic field of \qty{-1}{\tesla} is applied to prepare the flake in the $\downarrow \uparrow \downarrow$ state and we sweep the field to $B_{\text{exp}} = \qty{-2}{\milli\tesla}$ for imaging (first tile of \cref{fig:fig4}b). Subsequently, the bias magnetic field is increased to $B_{\rm{max}}$ in steps of $\qty{1}{\milli\tesla}$. After each field step, the field is swept back to $\qty{-2}{\milli\tesla}$ and the magnetization is measured again. When $B_{\rm{max}}$ reaches  $B_{\rm{co}}$, the magnetization reverses to the $\uparrow \downarrow \uparrow$ state and remains there for larger $B_{\rm{max}}$ (second tile of \cref{fig:fig4}b). Because the measurement field $B_{\text{exp}}$ is sufficiently small, it does not switch the flake back to the $\downarrow \uparrow \downarrow$ state (third tile of \ref{fig:fig4}b). We repeat this procedure at different temperatures.

The coercive fields of both flakes are shown in \cref{fig:fig4}c. The magnitudes of the coercive fields lead to the conclusion that magnetization reversal starts at nucleation sites, after which a domain wall propagates through the sample. If all spins in the flake were to reverse coherently, the coercive field would be given by the Stoner--Wohlfarth switching law \cite{stonerMechanismMagneticHysteresis1948}:
\begin{equation}
    B_{\text{co}} = \frac{2 E_K N}{\mu_\text{B}gS}(\sin^{2/3}\theta + \cos^{2/3}\theta)^{-3/2},
\end{equation}
where $E_K$ is the energy associated with the intrinsic magnetic anisotropy, $N$ is the number of layers, $M_\text{s}$ is the saturation magnetization, and $\theta = \pi - \phi$ where $\phi$ is the angle between the applied magnetic field and the magnetization. Using the bulk anisotropy parameter $E_K = \qty{0.0058}{\milli\electronvolt}$ \cite{calderMagneticStructureExchange2020} and $\theta = \qty{20}{\degree}$, this results in $B_{\text{co}}^{\text{SW}} = \qty{0.12}{\tesla}$ (at \qty{0}{\kelvin}). If the DFT-based trilayer value $E_K = \qty{0.03}{\milli\electronvolt}$ for the anisotropy is used, we find $B_{\text{co}}^{\text{SW}} = \qty{0.6}{\tesla}$, which roughly corresponds to the coercive fields seen in \cref{fig:fig3}d.  Since both of these values are much larger than the measured coercive fields, shown in \cref{fig:fig4}c, we conclude that magnetization reversal does not occur via coherent rotation. Instead, reversal is likely initiated by the nucleation of a small reversed domain, which requires a lower magnetic field than reversing the entire flake coherently. The associated energy barrier is expected to be somewhat lower than the energy required to form a closed circular domain wall around the reversed region, as nucleation is likely to occur at a pre-existing defect or sample edge. Once the domain has nucleated, the remainder of the region reverses through rapid domain-wall propagation on a timescale much shorter than that of our measurements. The behavior shown in \cref{fig:fig4}d,e suggests that domain-wall propagation is pinned by structural inhomogeneities. These inhomogeneities can arise from the transfer process, which inevitably introduces strain, or from local damage induced by the scanning tip, resulting in independently switching regions.

We describe the magnetization reversal of the flakes by considering the $\uparrow \downarrow \uparrow$ and $\downarrow \uparrow \downarrow$ configurations as two local minima of an energy landscape, separated by an energy barrier. During field reversal, the applied magnetic field and local dipolar fields progressively lower the activation barrier for reversal from the metastable state, while thermal fluctuations enable stochastic transitions to the stable state \cite{tremoletdelacheisserieMagnetismMaterialsApplications2005}. The transition rate is described by the Néel--Arrhenius law \cite{neel_theorie_1949},
\begin{equation}
    \Gamma = \Gamma_\text{0} e^{-\Delta/k_\text{B}T},
\end{equation}
where $\Gamma_\text{0}$ is the attempt frequency, $\Delta$ is the energy barrier, and $T$ is the temperature. Combining the Néel--Arrhenius law with a domain-nucleation model yields a coercive field that decreases linearly with temperature,
\begin{equation}
    B_{\text{co}} = B_0 - CT,
\end{equation}
where $B_0$ is the zero-temperature coercive field and $C$ depends on the nucleation-site properties. This behavior reproduces our observations, as shown by linear fits to the measured coercive fields in \cref{fig:fig4}c.

During the measurements, both flakes sustained damage due to tip puncture that is optically visible (Supporting Information \cref{fig:damagedflakes}, \cref{fig:coercivefieldsall}). Repeating the magnetization reversal measurements reveals that the coercive fields of flake 2 decrease substantially, with the magnitude decreasing 5 times at \qty{4.2}{\kelvin} after the damage. This observation suggests that the damage created new nucleation sites. A consistent trend is seen in flake 1, whose coercive fields also decrease at \qty{4.2}{\kelvin}, although by a smaller amount (\qty{25}{\percent}). Since the pristine coercive field of flake 1 was already considerably lower than that of flake 2, this suggests that the pre-existing nucleation sites in flake 1 had a lower nucleation barrier. The subsequent measurement results are shown in Supporting Information \cref{fig:coercivefieldsall}.

The appearance of magnetic domains is further supported by direct magnetic imaging. In \cref{fig:fig4}d,e we show the stray field of flake 1 at bias fields just below its coercive field. We observe that the magnetization reverses locally in parts of the flake while most of it remains in its original magnetic state. Such localized regions with magnetization differing from that of the surrounding material have previously been attributed to structural defects in van der Waals magnets \cite{thielProbingMagnetism2D2019}.

For a nucleation site with area $A$, the domain-wall energy scales as $A^{1/2}$, while the thermal activation term scales as $A^{-1}$, giving
\begin{equation}
    B_{\rm{co}} \propto CA^{-1/2} - TA^{-1},
\end{equation}
with $C$ a constant \cite{tremoletdelacheisserieMagnetismMaterialsApplications2005}. A derivation of this equation is given in Supporting Information \cref{sec:nuclSite}. Consequently, introducing additional nucleation sites or reducing the characteristic nucleation area is expected to decrease the low-temperature coercive field and increase the magnitude of the temperature dependence. The higher coercive field reported by \cite{wangConfigurableAntiferromagneticDomains2025} is consistent with a larger effective nucleation area or a higher domain-wall nucleation energy than in our flakes. The adjusting of coercive fields with temperature and nucleation site area adds another tunability factor to van der Waals antiferromagnets.

In summary, we have used scanning NV magnetometry to quantitatively probe the magnetization of trilayer CrPS$_4$ as a function of magnetic field and temperature. DFT-based Monte Carlo simulations reproduce the temperature dependence of the magnetization and show that coherent rotation cannot account for the measured coercive fields. Instead, the temperature dependence of the coercive field indicates that magnetization reversal proceeds via thermally activated domain nucleation following a Néel-Arrhenius law, which is further corroborated by magnetic imaging that reveals independently switching regions, likely pinned by structural inhomogeneities. Together, these observations provide a consistent picture of the magnetic behavior of this atomically thin antiferromagnet.

These results establish scanning NV magnetometry as a quantitative probe of both the equilibrium magnetic properties and the magnetization reversal of two-dimensional antiferromagnets, highlighting its exceptional sensitivity for addressing the rich magnetic phenomena emerging in atomically thin antiferromagnets, many of which remain inaccessible in the bulk counterparts. More broadly, our work provides insight into the microscopic mechanisms governing magnetization reversal in layered magnetic materials and opens new opportunities to investigate nanoscale spin dynamics in van der Waals magnets relevant to spintronic and magnonic technologies.

\section*{Supporting information}
The data underlying this study are openly available at Zenodo \href{https://doi.org/10.5281/zenodo.24165568}{https://doi.org/10.5281/zenodo.24165568}.

The Supporting Information is available free of charge at is available free of charge:
Methods, Optical images of sample damage, Coercive fields after damage occurred to the flake,  Mean-Field Theory for an Antiferromagnetic Trilayer, Magnetization along the \textit{a}-axis as function of bias magnetic field, DFT-computed in-plane magnetization, Scaling of the coercive field with the nucleation site area, Estimation of the magnetic moment per Cr atom (PDF)

\section*{Author Contributions}
Conceptualization: S.M.-V. and T.S.; Crystal growth and exfoliation and flake placement: S.M.-V.; Sample design and fabrication: R.W.M., S.M.-V., S.K.;  Constructing experimental setup: S.K., R.W.M., F.G., and T.S.; Measurements and software: R.W.M., S.M.-V., F.G., and P.C.J.A.Z.; Data analysis: R.W.M., F.G., S.M.-V. and T.S.; DFT (setup, simulations and writing): B.N and J.F.; Mean-field theory: F.G.; Landau expansion: F.M.; Figures: F.G., with the support of R.W.M. and S.M.-V.; Funding acquisition: S.M.-V. and T.S.; Project administration: S.M.-V. and T.S.; Supervision: S.M.-V. and T.S.; Writing – original draft: F.G. and S.M.-V.; Writing – review and editing: F.G., S.M.-V., R.W.M. and T.S. The manuscript was written through the contributions of all authors. All authors have given approval to the final version of the manuscript.

\section*{Acknowledgements}
This research was supported by The Kavli Foundation, by the Dutch Research Council (NWO) via ENW-XL grant OCENW.XL21.XL21.058 and VIDI grant VI.Vidi.193.077, and by the European Research Council via ERC CoG 101170480 MAGICWAVE. S.K. acknowledges support from NWO via VI.Veni.222.296 \href{https://doi.org/10.61686/OAAQA03904}{(https://doi.org/10.61686/OAAQA03904)}. B.N. and J.F. have been funded by\\
\noindent MCIN/AEI/10.13039/501100011033/FEDER, UE via projects PID2022-137078NB-100 and\\
\noindent PID2025-169100NB-I00, by the TRILMAX Horizon Europe consortium (Grant No. 101159646), and by Agencia SEKUENS (Asturias) under grant UONANO IDE/2024/000678 with the support of FEDER funds. S.M.-V. acknowledges the support from the Spanish MICIU (Ramón y Cajal program RYC2024-048264-I) and the Kavli Foundation (Synergy postdoctoral grant by the Kavli Institute of Nanoscience Delft).
\bigbreak
\noindent Present Addresses:
$^\#$S.K.: Heriot-Watt University, EH14 4AS, United Kingdom

\printbibliography
\end{refsection}

\appendix
\setcounter{secnumdepth}{2}
\begin{refsection}
\setcounter{figure}{0}
\setcounter{equation}{0}
\renewcommand{\theequation}{S\arabic{equation}}
\renewcommand{\theHequation}{S\arabic{equation}}
\renewcommand{\thefigure}{S\arabic{figure}}
\section*{Supporting information}
\startcontents[supplement]
\printcontents[supplement]{l}{1}{
\setcounter{tocdepth}{2}}
\pagenumbering{arabic}
\section{Methods}
\subsection{Sample preparation and NV setup}\label{sec:sampleprep}
\begin{figure}[hbt!]
    \centering
    \includegraphics[width=0.7\linewidth]{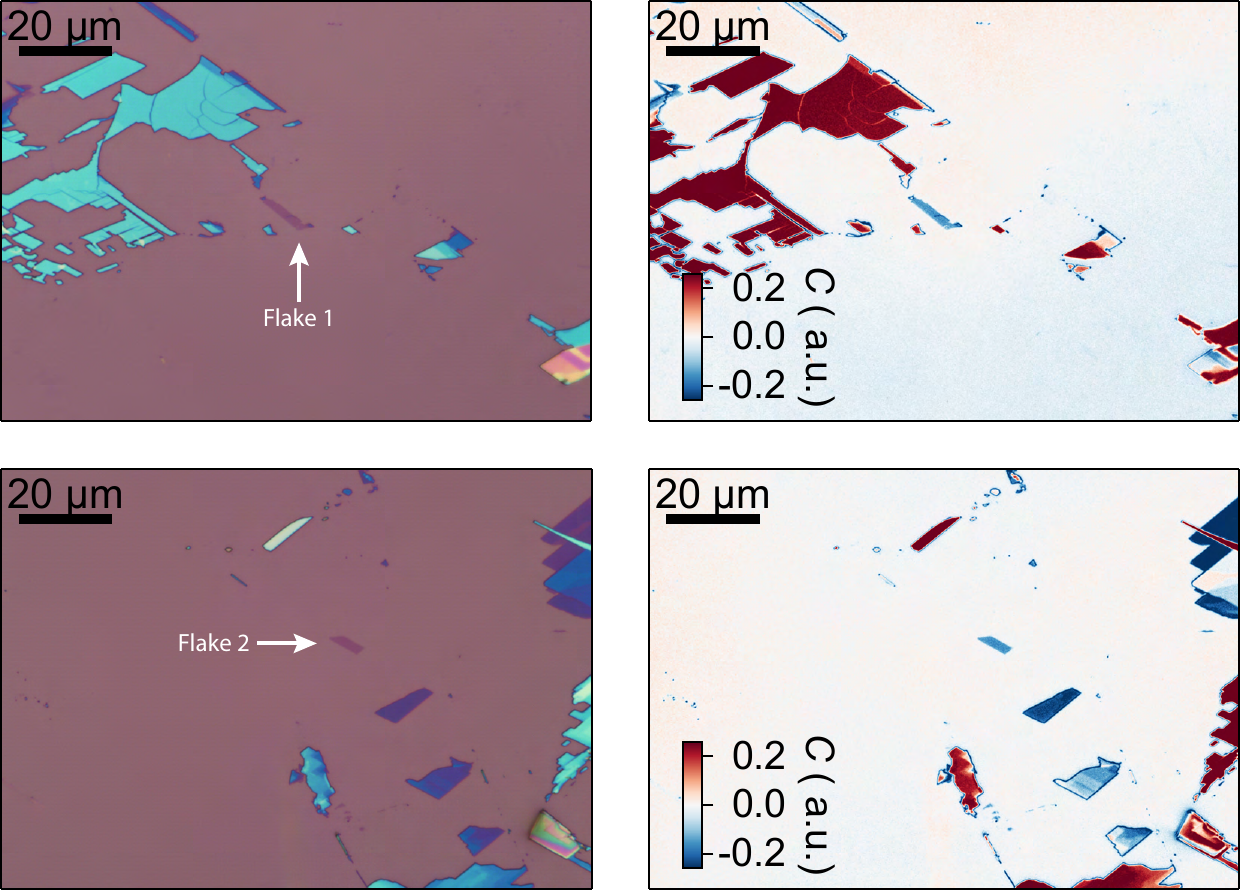}
    \caption{\textbf{Microscope images (left) and optical contrast images (right) of flakes 1 and 2.}}
    \label{fig:optContrast}
\end{figure}

Crystals of CrPS$_4$ were grown by chemical vapor transport using iodine as the transport agent and characterized structurally, compositionally, and magnetically, as previously reported by our groups in \cite{houmesHighlyAnisotropicMechanical2024,ghiasiQuantumSpinHall2025}. The number of layers of the exfoliated CrPS$_4$ was determined by evaluating the optical contrast \cite{leeStructuralOpticalProperties2017} and corroborated by NV magnetometry. The microscope images and optical contrast images are shown in \cref{fig:optContrast}.
The exfoliated flakes were stamped onto a coplanar waveguide fabricated by electron beam lithography (\qty{10}{\micro\meter} width; \qty{100}{\nano\meter} tickness Au) and covered with a layer of $\sim 20$ nm h-BN on top by using a deterministic transfer process using PDMS/PC \cite{wangOneDimensionalElectricalContact2013}. The sample and NV tip (a $\braket{111}$ QZabre scanning NV tip) are placed onto positioning stacks, as shown in Supplement \cref{fig:stacks}. The stacks are located inside an AttoLIQUID 1000 cryostat enabling measurements in the temperature range of \qtyrange{4.2}{100}{\kelvin} (approx.). The cryostat is equipped with an out-of-plane magnet that enables magnetic fields up to \qty{1}{\tesla}. Optical excitation of the NV center is provided by a \qty{514}{\nano\metre} laser (Coherent OBIS). The resulting photoluminescence is filtered using a dichroic mirror and a band-pass filter with a transmission window from 664 to 842 nm, after which the signal is detected by an avalanche photodiode. Mechanical excitation and readout of the AFM tip is performed by a Zurich Instruments MFLI. During a line scan, we employ the method described in fig. 1c to measure the magnetic field. Details of this method are described below.
\begin{figure}[hbt!]
    \centering
    \includegraphics[width=0.45\linewidth]{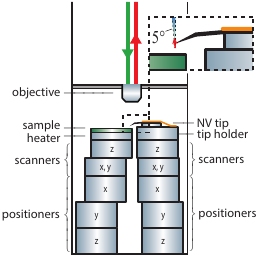}
    \caption{\textbf{Schematic of the sample and tip stacks in our measurement setup.} The NV tip, indicated by the red arrow, is mounted at an angle of 5\textdegree\ with respect to the sample normal.}
    \label{fig:stacks}
\end{figure}

\subsection{Bimodal feedback algorithm}\label{sec:bimodelreference}
Line scans are conducted by first performing a touch-down of the tip onto the sample on the starting location of the scan to find the touchdown height $z_0$. Then, we move the sample to perform a touchdown at the end location of the line scan. This allows us to account for sample tilt. When performing 2D scans, the tilt is determined based on $z_0$ of the previous line scan. We perform line scans at a constant lift height with respect to the sample, adjusting for its tilt.

In order to execute fast line scans, we use a bimodal feedback algorithm \cite{navaantonioMagneticImagingStatistical2021}. We start our measurements at the first pixel by performing a full ESR sweep (top part of fig. 1c). This measurement gives us the center frequency $\bar{\nu}_{\text{ESR}}$, contrast $c_{\text{ESR}}$, and full-width half-maximum (FWHM). The window size $\bar{\nu}_+ - \bar{\nu}_-$ is equal to the FWHM of the ESR signal for optimal sensitivity.

The error $D$ (bottom part of fig. 1c) is derived by modeling the ESR spectrum as a Lorentzian and Taylor expanding $D(\bar{\nu}) = PL(\nu_+) - PL(\nu_-)$ about the estimate $\bar{\nu} = \nu_{\text{ESR}}$ to first order. When there is a detuning $\bar{\nu}$ with respect to the ESR center frequency $\bar{\nu}_{\text{ESR}}$, $D(\bar{\nu})$ will be finite. The window is then shifted by a number given by the slope of the linear part around $\bar{\nu}_{\text{ESR}}$. This way, a changing $\bar{\nu}_{\text{ESR}}$ can be tracked by applying only two alternating microwave pulses with frequencies $\nu_-$ and $\nu_+$ respectively.

\begin{figure}[hbt]
    \centering
    \includegraphics[width=0.6\linewidth]{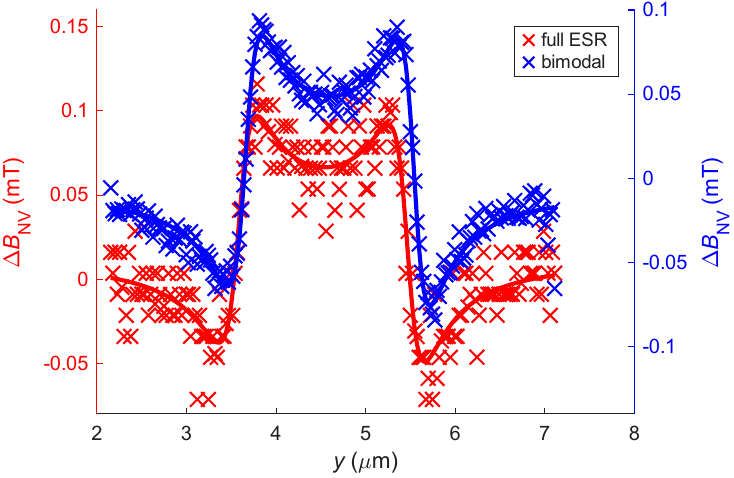}
    \caption{\textbf{Comparison of bimodal algorithm and the reference method of performing a touchdown and ESR measurement at each point.} Line scan is taken over a trilayer of CrPS$_4$ at a set lift of $h = \qty{120}{\nano\metre}$, an external field of $B_z = \qty{510}{\milli\tesla}$, and with a temperature of $T = \qty{4.2}{\kelvin}$. For clarity, a vertical offset is applied between the data in the plotting. Fits containing two odd Lorentzians, one for each edge of the flake, are included as the solid lines.}
    \label{fig:ESRcomparison}
\end{figure}

To test the accuracy of the algorithm, we have performed two separate line scans over the same region of a trilayer CrPS$_4$ flake using the bimodal scanning algorithm, and a reference measurement. In the reference measurement, a touchdown is performed at each measurement point, after which a full electron spin resonance (ESR) measurement is executed. The results are shown in figure \ref{fig:ESRcomparison}. Fitting both curves yields magnetizations of \(\qty{0.138(6)}{\milli\ampere}\) and \(\qty{0.136(4)}{\milli\ampere}\) for the bimodal and reference methods, respectively. The \(1\sigma\) uncertainties were estimated using Monte Carlo propagation of the uncertainties in the ESR resonance frequencies. We attribute small differences in the locations of the curves to thermal drifts.

\subsection{Fitting Procedure}\label{sec:fitting}
\subsubsection{Magnetization and stray field model}
In order to extract the magnetization from our line scans we fit our magnetic field data to a model considering homogeneous magnetization of the flake. In the case of a two-dimensional homogeneously magnetized film in the $xy$-plane filling the half space $x>0$, the magnetic field is given by \cite{tschudinImagingNanomagnetismMagnetic2024}:
\begin{equation}
    \mathbf{B}(\mathbf{r}) \approx \frac{\mu_0}{2\pi r}\left( M_z(\hat{r}\times\hat{y}) + M_x\hat{r}\right),
    \label{eq:fieldprofile}
\end{equation}
with $\mu_0$ the vacuum permeability. Thus, the magnetic field decays as $1/r$, where $r$ is the distance from the edge. When scanning across the edge at a constant lift height, the measured out-of-plane magnetic field $B_z(x)$ exhibits an odd-Lorentzian profile for purely out-of-plane magnetization and a Lorentzian profile for purely in-plane magnetization, with both profiles centered at the edge. When scanning across the width of our trilayer flakes, the measured field is a combination of the contributions from the two edges, each of which produces a field profile described by \cref{eq:fieldprofile}, leading to
\begin{equation}\label{eq:fittingEq}
    \mathbf{B}(\mathbf{r}) \approx \mathbf{B}_{\text{bias}}+ \sum_{i=1,2}\frac{\mu_0}{2\pi r_i(\mathbf{r})}\left[(\mathbf{M}\cdot\hat{\mathbf{n}}_{\text{sample}})(\hat{r}_i(\mathbf{r})\times\hat{\mathbf{l}}_{\text{edge},i}) + (\mathbf{M}\cdot\hat{\mathbf{n}}_{\text{edge},i})\hat{r}_i(\mathbf{r})\right]
\end{equation}
Here, we introduce an arbitrary reference frame by defining $\hat{\mathbf{n}}_{\text{sample}}$ as the normal to the sample plane, and $\hat{\mathbf{l}}_{\text{edge},i}$ and $\hat{\mathbf{n}}_{\text{edge},i}$ as the unit vectors parallel and normal to edge $i$, respectively. We define the vector
\begin{equation}
    \mathbf{r}_i(\mathbf{r})
    =
    \mathbf{r}-\mathbf{r}_{\text{edge},i}
    -
    \left[
        \left(\mathbf{r}-\mathbf{r}_{\text{edge},i}\right)
        \cdot\hat{\mathbf{l}}_i
    \right]\hat{\mathbf{l}}_i
\end{equation}
as the shortest vector connecting the point $\mathbf{r}$ to edge $i$, where $\mathbf{r}_{\text{edge},i}$ is an arbitrary point on the edge. 

The expression above describes the full magnetic-field vector. However,
the NV center measures only the component of the magnetic field along
its axis, $\hat{\mathbf{e}}_{\text{NV}}$. The detected field is therefore given by
\begin{equation}
    B_{\text{NV}}(\mathbf{r})
    =
    \mathbf{B}(\mathbf{r})\cdot\hat{\mathbf{e}}_{\text{NV}}.
\end{equation}
We use this projected field as the quantity to fit to the measured line scans.

\subsubsection{Model parameters and assumptions}
For the procedure of fitting the line scans to equation \cref{eq:fittingEq}, we use a reference frame in which the $z$-axis is approximately normal to the sample plane, while the $x$- and $y$-axes are oriented parallel and perpendicular to the coplanar waveguide, respectively.

We assume that any in-plane magnetization component lays along the intermediate axis ($a$) of the crystal, which is shown for both flakes in \cref{fig:fig1}b and \cref{fig:fig4}a.

Owing to a small tilt between the sample normal and the $z$-axis, we compensate for the sample inclination to maintain a constant sample--sensor standoff distance of approximately \qty{100}{\nano\meter} during each line scan. Prior to each line scan, we perform a touchdown at the start ($x_1$) and end ($x_2$) positions of the scan, yielding the corresponding contact $z$-scanner voltages $V_{z,\text{contact}}(x_1)$ and $V_{z,\text{contact}}(x_2)$. We then define the scanning plane for that line by linearly interpolating between these two contact positions and adding a constant lift voltage:
\begin{equation}
    V_{z,\parallel}(x) = -V_{\text{lift}}
    + V_{z,\text{contact}}(x_1) +
    \frac{V_{z,\text{contact}}(x_2) - V_{z,\text{contact}}(x_1)}
    {x_2-x_1} (x-x_1).
\end{equation}
This procedure is repeated independently for every line scan, such that the scanning plane is redetermined from the local sample topography before each line scan. We find from our fits that the sample tilt for both samples is below \qty{5}{\degree}.

\subsection{DFT methods and parameters}\label{sec:DFTsup}
The electronic structure of the systems was calculated using the SIESTA density functional theory (DFT) code \cite{Soler2002,cuadradoValidityOnsiteSpinorbit2021}.
The trilayer system was modeled using a unit cell containing 4 Cr, 4 P, and 16 S atoms in each of the three layers. 
The in-plane lattice constants were chosen as $a=\qty{10.91}{\angstrom}$, and $b=\qty{7.36}{\angstrom}$, based on our monolayer calculations \cite{nagyfalusiMagneticGroundStates2026}, while the out-of-plane lattice parameter and monoclinic angle were set to $c=\qty{6.1382}{\angstrom}$ and $\beta=\qty{91.886}{\degree}$, respectively, following the experimental values reported in \cite{Fas2025}.  
The magnetic moment per layer was found to be $2.943\,\mu_\mathrm{B}$.

To describe the magnetic interactions, we employed different approaches for the interlayer and intralayer interactions. 
The interlayer exchange interactions were calculated using the GROGU code \cite{Grogu2023}, considering the first three interacting shells, as illustrated in \cref{fig:fig3}a. Note that there are two subshells of first layer interactions due to the dimerisation along the $b$ direction.
For each layer $p$, the spin Hamiltonian is given by
\begin{align}
H_{p}&=
\frac{1}{2}\sum_{i,j} \mathbf{e}_{p,i}
\mathbf{J}_{pi,pj}
\mathbf{e}_{p,j}
+ \sum_{i} A^{xx}(\mathbf{e}_{p,i}^x)^2
+ \sum_{i} A^{yy}(\mathbf{e}_{p,i}^y)^2
- \sum_{i}\mu_p \mathbf{e}_{p,i} \cdot \mathbf{B},
\label{eq:heis_mod}
\end{align}
where $\mathbf{J}_{pi,pj}$ is a $3\times3$ tensor describing the exchange interaction between sites $i$ and $j$ in layer $p$. 
The quantities $A^{xx}$ and $A^{yy}$ describe the on-site anisotropy interactions along the $x$ and $y$ directions, respectively. 
The last term represents the Zeeman interaction, where $\mu_p$ is the magnitude of the magnetic moment and $\mathbf{B}$ is the external magnetic field. 
Finally, $\mathbf{e}_{p,i}$ is a unit vector associated with site $i$ in layer $p$, pointing along the local magnetization direction.

Since the van der Waals interaction between the layers is difficult to describe accurately within the DFT calculations, the effective interlayer interaction, $J_0^{\mathrm{IL}}$, was treated as a fitting parameter. 
Its value was determined by fitting the spin model to the experimental magnetization curves presented in this work.

The total Hamiltonian is given by
\begin{align}
     H&= \sum_{p=1}^3 H_p + \sum_{\left\langle p,q\right\rangle} \sum_i J_0^\mathrm{IL} (\mathbf{e}_{p,i} \cdot\mathbf{e}_{q,i})
      \,, 
    \label{eq:heis_mod_full}
\end{align}
where the interlayer coupling acts only between neighboring layers.

The isotropic exchange interactions, averaged over equivalent bonds within each layer, are listed in \cref{tab:exchange}. 
The exchange parameters show little variation between layers, with the exception of $J_2$, which changes sign in the middle layer. 
The interlayer coupling parameter was treated as a fitting parameter, which identifies it to be $J_0^{\mathrm{IL}}=\qty{0.2}{\milli\electronvolt}$. 
The anisotropy constants were found to be identical in all three layers, with $A^{xx}=\qty{0.027}{\milli\electronvolt}$ and $A^{yy}=\qty{0.036}{\milli\electronvolt}$.

\begin{table}[hbt!]
 \caption{Isotropic exchange interaction in each layer  in meV units}
    \label{tab:exchange}
    \centering
    \begin{tabular}{c|ccc}
              &1 &2&3 \\ \hline
      $J_{1a}$  & 3.40 & -3.40  & -3.39     \\
      $J_{1b}$  & -10.51 &	-10.50	&-10.49  \\
       $J_{2}$  & -0.04 &	0.04 &	-0.04  \\
        $J_{3}$  & -1.68 &	-1.60  &-1.68  \\    
    \end{tabular}
\end{table}

The temperature dependence of the magnetization was modeled using Monte Carlo (MC) simulations. 
Each layer was represented by a $64\times64$ array of unit cells, with each unit cell containing four Cr sites, corresponding to a total of $3\times4\times64\times64=49\,152$ Cr sites for the trilayer system. 
The simulations were initialized with a random spin configuration and subsequently cooled with a temperature step of $\qty{1}{\kelvin}$. 
At each temperature, the system was equilibrated for 10\,000 MC steps between successive measurements, with 150 measurements used to calculate the thermal average. 
The heat-capacity curves yield a critical temperature of approximately $\qty{40}{\kelvin}$, within the range of the experimentally measured values. 
The calculated magnetization curves are presented in \cref{fig:fig3}b.

For a better understanding of the magnetic behavior, we consider the layer-resolved magnetization curves shown in \cref{fig:fig3}c for $B=0$ and $\qty{1}{\tesla}$. 
In zero field, the magnetization of each layer averages to zero above the critical temperature, while below $T_c$ the antiferromagnetic (AFM) ordering develops immediately. 
In the presence of an external field, the magnetic moments become aligned already above $T_c$, resulting in a region where all three layers are ferromagnetically aligned. 
Upon further cooling, the magnetization of the middle layer changes sign, while the top and bottom layers remain aligned with each other, leading to a decrease in the total magnetization. 
With increasing field, this reversal occurs at lower temperatures, extending the ferromagnetic region and producing a more pronounced local maximum in the $M(T)$ curve.

\subsection{Coercive field measurement methods}
The coercive fields of flake 1 were determined by increasing the bias magnetic field in steps between 2.5 and \qty{5}{\milli\tesla} for the pristine flake, or \qty{1}{\milli\tesla} after tip damaging, and measuring the stray field of the flake at this bias field by acquiring a line scan. The field is swept upward and downward after an initial field saturation at -1 T. The hysteresis loops are symmetric, indicating the absence of an exchange-bias-induced field offset.

\begin{figure}[hbt!]
    \centering
    \includegraphics[width=0.5\linewidth]{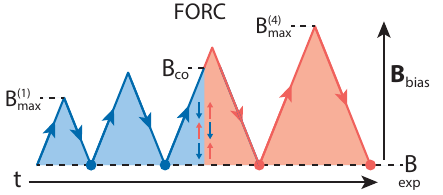}
    \caption{\textbf{Measurement procedure for finding the coercive fields of flake 2.} A first-order reversal curve (FORC) protocol was used, with $B_{\text{exp}} = $ \qty{-2}{\milli\tesla} in the $\hat{z}$-direction.}
    \label{fig:coercMethodsFlake2}
\end{figure}

For flake 2, due to a limited ODMR contrast at many relevant fields, we opted to use a FORC protocol, as shown in \cref{fig:coercMethodsFlake2}. The step size in these measurements was between 0.5 and \qty{1}{\milli\tesla} for the pristine flake, or \qty{0.5}{\milli\tesla} for the damaged flake. $B_{\rm{co}}$ is determined, within an accuracy set by the step size, as the first $B_{\text{max}}$ that switches the flake to the $\uparrow\downarrow\uparrow$ state. 

\clearpage
\section{Optical images of pristine and damaged trilayers}
\begin{figure}[hbt!]
    \centering
    \includegraphics[width=\linewidth]{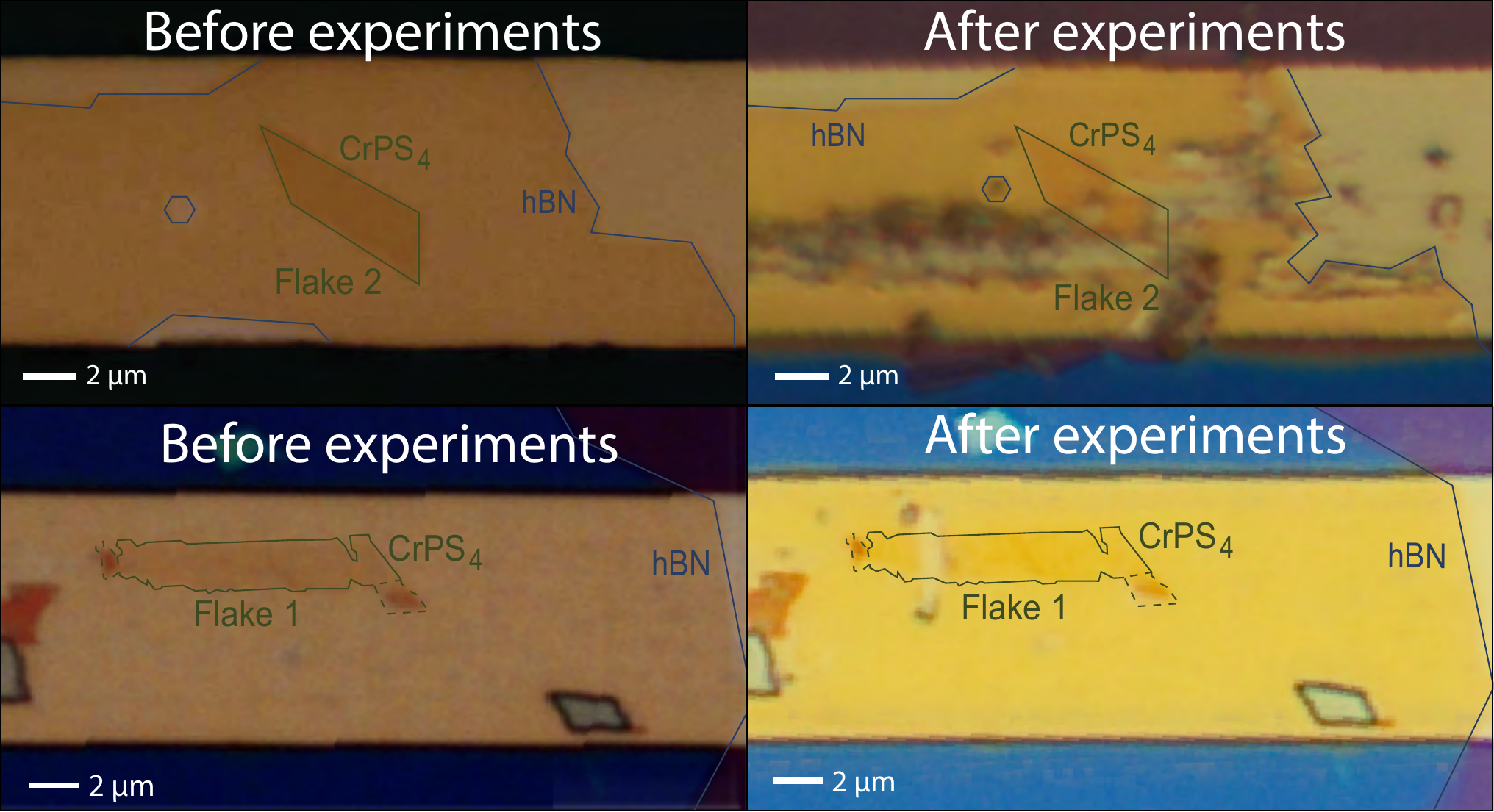}
    \caption{\textbf{Images showing damage to both flakes.} Optical images showing pristine (left) and damaged (right) flakes. The damages occurred at some moment during the measurements. Flakes are indicated by a green outline, and they are encapsulated by a layer of $\sim\qty{20}{\nano\metre}$ h-BN, which is blue outlined.}
    \label{fig:damagedflakes}
\end{figure}

\clearpage
\section{Coercive fields in damaged trilayers}
\begin{figure}[hbt!]
\centering
    \includegraphics[width=0.8\linewidth]{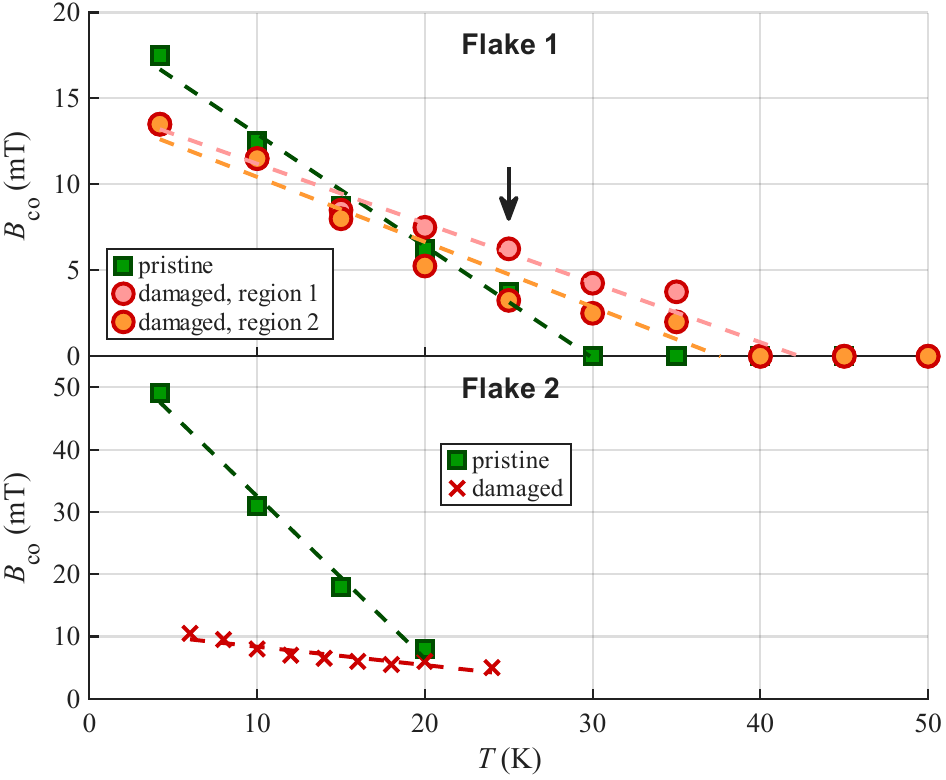}
    \caption{Coercive fields of both flakes before and after physical damage. For flake 1 after damage, two independently switching regions are treated as separate data sets. The dashed lines show linear fits to the data. Before damage, line scans were acquired at the location indicated in fig. 1d. After damage, they were acquired along the dashed line in fig. 3c. The arrow indicates the data point for which the line scans in figure fig. 3e are shown.}
    \label{fig:coercivefieldsall}
\end{figure}

The coercive fields of both flakes, before and after damage, are shown in \cref{fig:coercivefieldsall}. For flake 1, since the region where the line scans were acquired after damage occurred contained two regions that switch separately, they are treated as independent data sets, indicated as \textcolor{darkgray}{damaged, region 1} and \textcolor{darkgray}{damaged, region 2}.

\clearpage
\section{Mean-Field Theory for an Antiferromagnetic Trilayer}\label{sec:MFT}
In order to derive a Mean-Field Theory model for a trilayer antiferromagnet, we use the following spin Hamiltonian:
\begin{equation}
    \mathcal{H} = -J_{\parallel} \sum_{\langle ij\rangle,l} \mathbf{S}_{i,l} \cdot \mathbf{S}_{j,l} -J_{\perp} \sum_{i,\langle l,l'\rangle} \mathbf{S}_{i,l} \cdot \mathbf{S}_{i,l'} - g \mu_{\text{B}}B \sum_{i,l}S_{i.l}^z,
\end{equation}
where $J_{\parallel}$ and $J_{\perp}$ are the intra- and interlayer exchange constants respectively, $\mathbf{S}_{i,l}$ is the spin operator of the Cr ion at lattice site $i$ in layer $l$, and $B$ is the applied magnetic field. We choose the $\hat{z}$-axis as the easy axis of the magnet. We assume the trilayer obeys the symmetry
\begin{equation}
    m_1 = m_3 \equiv m_A, \hspace{1 cm} m_2 \equiv m_B.
\end{equation}
This symmetry is expected for an A-type antiferromagnetic trilayer with equivalent outer layers. The magnetization is proportional to the total mean spin, $M(T) \propto \bar{S}^z(T)$. Assuming collinear order along the easy axis, Mean-Field theory neglects spin fluctuations by replacing the neighboring spin operators by their thermal averages,
\[
\mathbf S_j \rightarrow \bar S_j^z\hat{\mathbf z}.
\]
Following \cite{serhaUnifiedTheoryMagnetization2025}, the effective fields are expressed in temperature units (energy divided by $k_\text{B}$). Defining
\begin{equation}
    \mathcal{J}_\perp \equiv J_\perp Z_\perp, \hspace{1 cm} \mathcal{J}_\parallel \equiv J_\parallel Z_\parallel
\end{equation}
where $Z$ denotes the coordination number, the effective fields acting on spins in the outer (A) and central (B) layers are therefore
\begin{equation}
    H_\text{A}^{\text{eff}} = \bar{S}^z_\text{A} \mathcal{J}_\parallel + \bar{S}^z_\text{B} \mathcal{J}_\perp + g \mu_\text{B}B / k_\text{B}, \hspace{1 cm} H_\text{B}^{\text{eff}} = \bar{S}^z_\text{B} \mathcal{J}_\parallel + 2\bar{S}^z_\text{A} \mathcal{J}_\perp + g \mu_\text{B}B / k_\text{B},
\end{equation}
These effective fields give rise to the coupled mean-field self-consistency equations \cite{landauStatisticalPhysics1938}
\begin{equation}
    \bar{S}^z_\text{A}(T) = b_\text{S} \bigg[ \frac{\bar{S}^z_\text{A}(T) \mathcal{J}_\parallel + \bar{S}^z_\text{B}(T) \mathcal{J}_\perp + g \mu_\text{B}B / k_\text{B}}{T} \bigg], \hspace{1 cm} \bar{S}^z_\text{B}(T) = b_\text{S}\bigg[ \frac{\bar{S}^z_\text{B}(T) \mathcal{J}_\parallel + 2\bar{S}^z_\text{A}(T) \mathcal{J}_\perp + g \mu_\text{B}B / k_\text{B}}{T} \bigg],
\end{equation}
where $b_\text{S}(x)$ is the Brillouin function given by
\begin{equation}
    b_\text{S}(x) = \bigg( S + \frac{1}{2} \bigg) \coth{\bigg[ \bigg(S + \frac{1}{2} \bigg) x \bigg]} - \frac{1}{2} \coth{\bigg[ \frac{x}{2} \bigg]}.
\end{equation}
The expectation value of the total spin is therefore
\begin{equation}
    S^z(T)=2\bar S_A^z(T)+\bar S_B^z(T).
\end{equation}

\clearpage
\section{Magnetization along the \textit{a}-axis as function of bias magnetic field}
\begin{figure}[hbt!]
    \centering
    \includegraphics[width=0.6\textwidth]{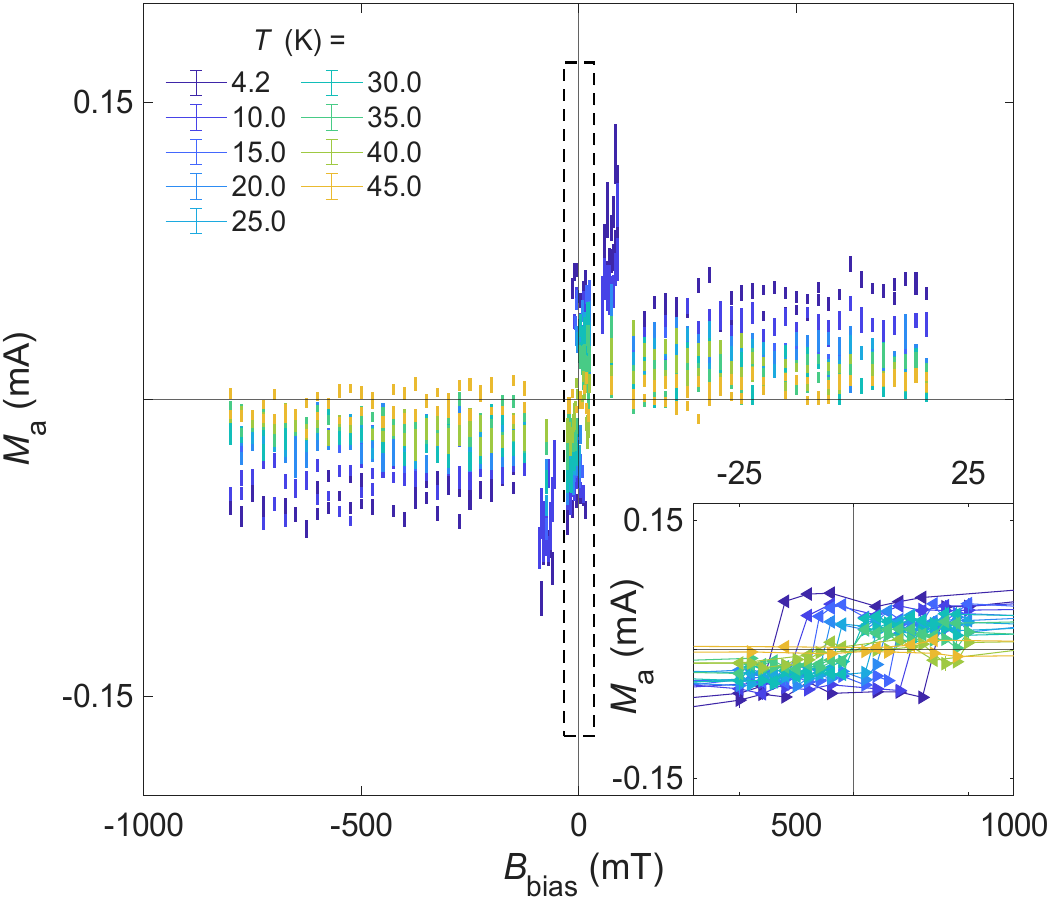}
    \caption{\textbf{Magnetization along the $a$-axis as a function of $B_{\text{bias}}$ during upward and downward field sweeps.} The inset shows a zoomed view of the low-field data, revealing a hysteresis loop. Triangles indicate the sweep direction.}
    \label{fig:magXvsB}
\end{figure}

The in-plane component of the magnetization shown in fig. 2b, assumed to lay along the ($a$) axis of the crystal, based on \cite{calderMagneticStructureExchange2020}, is shown in figure \cref{fig:magXvsB}.

\clearpage
\section{Scaling of the coercive field with the nucleation site area}\label{sec:nuclSite}
We consider magnetization reversal by thermally activated nucleation of a reversed magnetic domain in a two-dimensional ferromagnet. Let \(A\) denote the area of the nucleation site, \(\lambda\) the domain-wall energy per unit length, \(M_s\) the saturation magnetization per unit volume, and \(t\) the thickness of the magnetic layer. The domain-wall energy scales with \cite{tremoletdelacheisserieMagnetismMaterialsApplications2005}
\begin{equation}
    E_{\mathrm{DW}}=a\lambda A^{1/2},
\end{equation}
where $a$ is a dimensionless geometric factor determined by the shape of the nucleus. An applied magnetic field $B$, directed opposite to the initial magnetization, lowers the energy of the reversed state. For a nucleus of area $A$, the corresponding Zeeman energy gain is
\begin{equation}
    E_{\mathrm{Z}}= M_s BtA.
\end{equation}
We additionally account for a local dipolar field $B_{\mathrm{dip}}$, assumed to assist reversal. Within the nucleation model, the activation barrier is approximated by
\begin{equation}\label{eq:NA1}
    \Delta(B,A) = a\lambda A^{1/2} - M_stA(B+B_{\mathrm{dip}}).
\end{equation}
The switching rate is described by the Néel-Arrhenius law,
\begin{equation}
    \Gamma=\Gamma_0 \exp\!\left[-\frac{\Delta(B,A)}{k_{\mathrm B}T}\right],
\end{equation}
where $\Gamma_0$ is the attempt frequency. For a measurement with characteristic observation time $\tau_{\mathrm{obs}}$, switching becomes probable when $\Gamma\tau_{\mathrm{obs}}\sim1$. Consequently, the activation barrier at the coercive field $B_{\mathrm{co}}$ satisfies
\begin{equation}\label{eq:NA2}
    \Delta(B_{\mathrm{co}},A) = k_{\mathrm B}T \ln\!\left(\Gamma_0\tau_{\mathrm{obs}}\right).
\end{equation}
Substituting \cref{eq:NA2} into \cref{eq:NA1} and isolating $\Delta(B_{\mathrm{co}})$ yields

\begin{equation}
    B_{\mathrm{co}}
=
\frac{a\lambda}{M_st}A^{-1/2}
-B_{\mathrm{dip}}
-\frac{k_{\mathrm B}}{M_st}
\ln\!\left(\Gamma_0\tau_{\mathrm{obs}}\right)
TA^{-1}.
\end{equation}
Neglecting or absorbing the field-independent offset, the coercive field follows the scaling
\begin{equation}
    B_{\mathrm{co}}\propto A^{-1/2}-TA^{-1}.
\end{equation}

\section{Estimation of the magnetic moment per Cr atom}\label{sec:estCr}
The magnetic moment per Cr atom was estimated from the measured two-dimensional monolayer magnetization, $M_{\mathrm{2D}}$, using the crystallographic unit-cell area of CrPS$_4$. The two-dimensional magnetization is related to the magnetic moment per Cr atom by
\begin{equation}
    M_{\mathrm{2D}} = \frac{N_{\mathrm{Cr}}\mu_{\mathrm{Cr}}\mu_{\mathrm{B}}}{A_{\mathrm{uc}}},
\end{equation}
where $N_{\mathrm{Cr}}=4$ \cite{calderMagneticStructureExchange2020} is the number of Cr atoms per unit cell, $\mu_{\mathrm{Cr}}$ is the magnetic moment per Cr atom expressed in units of the Bohr magneton ($\mu_{\mathrm{B}}$), and $A_{\mathrm{uc}}=ab$ is the in-plane unit-cell area. Here, $a=\qty{1.08510(3)}{\nano\meter}$ and $b=\qty{0.72560(2)}{\nano\meter}$ are the in-plane lattice parameters \cite{calderMagneticStructureExchange2020}, giving $A_{\mathrm{uc}}=\qty{0.7881}{\nano\meter\squared}$.

Rearranging the expression gives
\begin{equation}
    \mu_{\mathrm{Cr}} =
    \frac{M_{\mathrm{2D}} A_{\mathrm{uc}}}
    {N_{\mathrm{Cr}}\mu_{\mathrm{B}}}.
\end{equation}
Using the experimentally determined two-dimensional magnetization of $M_{\mathrm{2D}}=\qty{0.136}{\milli\ampere}$ (\cref{fig:fig1}) and the Bohr magneton $\mu_{\mathrm{B}}=\qty{9.27e-24}{\ampere\meter\squared}$, we obtain
\begin{equation}
    \mu_{\mathrm{Cr}}\approx 2.89\ \mu_{\mathrm{B}}/\mathrm{Cr}.
\end{equation}
This value agrees with other experimental observations \cite{ghiasiQuantumSpinHall2025}.
\printbibliography
\end{refsection}

\end{document}